\documentclass[11pt,a4paper]{amsart}

\usepackage[latin1]{inputenc}
\usepackage[english]{babel}
\usepackage{amsmath}

\usepackage{amssymb, mathabx}
\usepackage[numbers]{natbib}
\usepackage{graphicx}
\usepackage{braket}
\usepackage[pdftex,plainpages=false,colorlinks,hyperindex,bookmarksopen,linkcolor=red,citecolor=blue,urlcolor=blue]{hyperref}

\usepackage{epstopdf}

\usepackage{braket}

\bibpunct{[}{]}{;}{n}{,}{,}
\usepackage[hmargin=3cm,vmargin={3.5cm,4cm}]{geometry}

\theoremstyle{definition}                                 %stile corsivo
\theoremstyle{definition}                           %stile roman
\theoremstyle{remark}                             %stile per osservazioni
\usepackage{color}

\def\eg{{\it e.g. }} 
\def\ie{{\it i.e. }}

\newcommand{\res}{\mathcal{R}es \,}

\newcommand{\blue}[1]{{\color{black}#1}}

\def\ds{\displaystyle}
\def\ds{\displaystyle}
\def\eg{{\it e.g.}\ }
\def\ie{{\it i.e.}\ }

\def\G{{\mathcal{G}}}
\def\J{{\mathcal{J}}}
\numberwithin{equation}{section}

\allowdisplaybreaks

\begin{document}
\title{A class  of linear  viscoelastic models	\\ 
based on Bessel functions}

    \author{Ivano Colombaro$^1$}
		\address{${}^1$ Department of Physics $\&$ Astronomy, University of 	
    	    Bologna and INFN. Via Irnerio 46, Bologna, ITALY.}
		\email{ivano.colombaro@bo.infn.it}
	
	    \author{Andrea Giusti$^2$}
		\address{${}^2$ Department of Physics $\&$ Astronomy, University of 	
    	    Bologna and INFN. Via Irnerio 46, Bologna, ITALY.}
		\email{andrea.giusti@bo.infn.it}
	
    \author{Francesco Mainardi$^3$}
    	    \address{${}^3$ Department of Physics $\&$ Astronomy, University of 	
    	    Bologna and INFN. Via Irnerio 46, Bologna, ITALY.}
			\email{francesco.mainardi@bo.infn.it}


    \date  {\today}%%{January 2016}

\begin{abstract}
In this paper we investigate a general class of linear viscoelastic models whose creep and relaxation memory  functions are expressed in Laplace domain by suitable ratios of modified Bessel functions of contiguous order.
In time domain  these functions are shown  to be expressed
by Dirichlet series (that is  infinite Prony series). It follows that 
the corresponding creep compliance and relaxation modulus turn out to be characterized by  infinite discrete spectra of retardation and relaxation time respectively.
As a matter of fact, we get a class of viscoelastic models depending on a real parameter $\nu > -1$.  Such models exhibit rheological properties
akin to those of a fractional Maxwell model (of order $1/2$) for short times and of a standard Maxwell model for long times.
	
\end{abstract}

    \maketitle
    
\section{Introduction} \label{section_intro}
    The purpose of this paper is to introduce   a rheological characterization of  a peculiar class of models generalizing with  infinite discrete spectra of retardation and relaxation times the classical Maxwell body. The corresponding  creep and relaxation memory  functions are expressed in Laplace domain as suitable ratios of modified Bessel functions  of contiguous order $\nu > -1$. In time domain  these functions are shown  to be expressed by Dirichlet series  (that is  infinite Prony series) so that 
the corresponding creep compliance and relaxation modulus turn out to be monotonic functions.  These functions are characterized by  infinite discrete spectra of characteristic times related to the zeros of  Bessel functions of the first kind.
  
    The plan of paper is as following. 
In  Section 2 we revisit the  essential notions of the linear theory of uni-axial viscoelasticity. The survey  will be carried  out by adopting the representations of the basic stress - strain relations  in the Laplace transform domain, as it is common  in linear viscoelasticity.
 In particular,  we will outline  the role of the so-called memory functions to characterize  the rate   of   creep and relaxation by using the mathematical notions  of completely monotonic functions.  
 
	 After this introductory  discussion,  in Section 3  we analyse a general  class of fluid-like models (depending on a parameter $\nu>-1$) that provides an interesting generalization of the classical Maxwell model.
 For this purpose
we  provide the expressions of the Laplace transforms  of the memory functions as ratios between of two modified Bessel functions of contiguous orders.
The memory functions are  then  shown to be  complete monotonic functions of time, expressed in terms of  Dirichlet series. 
Finally, by a suitable time integration, we derive the corresponding creep compliance and the relaxation modulus (the so-called material functions), so providing a full characterization of the whole class of our viscoelastic models. Due to the relation of these viscoelastic models with the Bessel  functions ,  we agree to refer them to as {\it Bessel models}.

In Section 4,  for the Bessel models we compute
 the asymptotic representations of the memory functions  and of the corresponding material functions  for short and long times. We show 
some instructive plots of the numerical results in order to outline the time range of validity of the asymptotic  representations. 
 We observe that the  creep and relaxation properties of our models are consistent with those of a standard Maxwell model for large times.   On the other hand, these models are compatible with a fractional Maxwell model of order 1/2, for short times.    

Finally, in Section 5  we complete the paper with concluding remarks and hints for future research.

\section{Generalities on linear viscoelasticity} \label{section_2}

According to the linear theory, a viscoelastic body
can be considered as a linear system
with the stress (or strain) as the excitation function (input)
and the strain (or stress) as the response function (output).
 In this respect, the response functions to
an excitation expressed by the Heaviside step function $\Theta(t)$
are known  to   play a fundamental role both from a mathematical
and physical point of view. We denote by $\mathcal{J}(t)$ the strain response
to the unit step of  stress,  according to the {\it creep test}
 and by $\mathcal{G}(t)$ the stress response to a unit step of strain,
according to the {\it relaxation test}.

The functions $\J(t)$ and $\G(t)$  are usually referred to as the
{\it creep compliance} and {\it relaxation modulus}
respectively, or, simply, the {\it material functions}
of the viscoelastic body.   In view of the causality
requirement, both  functions are    vanishing
for $t<0$.
%%%%%%%%

The limiting values of the material functions
for $t \to 0^+$ and $t \to +\infty$  are related to the
instantaneous (or glass) and equilibrium behaviours of the viscoelastic
body, respectively. As a consequence, it is usual to denote
   $\J_g := \J(0^+) \ge 0$  the {\it glass compliance},
   $ \J_e := \J(+\infty)\ge 0 $ the {\it equilibrium  compliance},
and
  $ \G_g := \G(0^+)\ge 0$  the {\it glass modulus}
  $ \G_e := \G(+\infty)\ge 0$   the {\it equilibrium modulus}.
As a matter of fact, both the material functions are
non--negative with 
 $ \J(t)$   {\it non decreasing} and
 $\G(t)$  {\it non increasing}.

Denoting by $\sigma(t)$ and $\epsilon(t)$ the uniaxial stress and strain, respectively, and 
under the hypotheses of sufficiently well behaved causal histories,
 in most cases the  constitutive equations can be written in the following forms
\begin{equation} \label{stress-strain}
\left\{
\begin{array}{ll}
{\ds \epsilon (t)}
&
= {\ds \J_g\, \sigma(t) + 
\int_0^t  \!\! \dot \J(t-t')\,\sigma (t' ) \, dt' }\,,
  \\  \
{\ds \sigma  (t)}
&
=  {\ds \G_g\, \epsilon (t) + \int_0^t  \!\! \dot \G(t-t')\,
 \epsilon  (t') \, dt'}\,.
 \end{array}
 \right.
 \end{equation}
Being of convolution type,  equations (\ref{stress-strain}) can be conveniently
treated by the technique of Laplace transforms so they read
in the Laplace domain
\begin{equation}
\widetilde \epsilon (s) = s\, \widetilde{\J}(s) \, \widetilde \sigma(s)\,,
\quad
\widetilde \sigma  (s) = s\, \widetilde{\G}(s) \, \widetilde \epsilon (s)\,,
\end{equation}
from which we derive the {\it reciprocity relation}
\begin{equation} \label{reciprocity}
  s\, \widetilde{\J}(s)  = \frac{1}{s\,\widetilde{\G}(s)}
 \,. 
 \end{equation}
Due to  limiting theorems for the Laplace transform, we deduce that
\begin{equation} \label{limits}
\J_g = {1}/ {\G_g}, \quad  \J_e = {1}/{\G_e}\,, 
\end{equation}
with the convention that $0$ and $+\infty$ are reciprocal to each other.

 The causal functions $\dot \J(t)$  and $\dot \G(t)$ are referred  as
the {\it rate of creep (compliance)} and
 the {\it rate of relaxation (modulus)}, respectively;
they play the role of {\it memory functions}  in the constitutive
equations (\ref{stress-strain}).
  
Assuming  $\J_g >0$  and  $\G_g>0$, that is restricting our attention 
to viscoelastic models exhibiting instantaneous elasticity  
 it is convenient
to consider   the memory functions
scaled with their corresponding  initial   values:
\begin{equation} \label{Psi-Phi} 
\Psi(t) =  \frac{1}{\J(0^+)}\,  \frac{d\J}{dt}\,, \quad
\Phi(t) = -\frac{1}{\G(0^+)}\, \frac{d\G}{dt}\,.
\end{equation}
These functions are here required to be  completely monotonic (CM) functions, that is they are  non--negative, non-increasing  functions for $t>0$ with infinitely many derivatives alternating   in sign.
As outlined by several authors, see e.g. \cite{Hanyga STAMM04},
\cite{Mainardi_BOOK10}, 
the condition of completely monotonicity  is sufficient for the physical acceptability and realizability of the models since it ensures, for instance, that in isolated systems the energy decays monotonically as expected from physical considerations. 
For mathematical details on CM functions  we refer the interested reader to the excellent treatise \cite{Schilling-et-al 2012}.

%% by means of  discrete or continuous distributions 
%% of relaxation  elements. 
Then 
\begin{equation} \label{J-G}
\begin{split}
 \J(t) &= \J(0^+)\left[1 + \int _0 ^t \Psi (t') \, dt'\right]\,,\\
  \G(t) &= \G(0^+)\left[1 - \int _0 ^t \Phi (t') \, dt'\right]\, .
\end{split}
\end{equation}
Because of the reciprocity relations (\ref{reciprocity}) and the above equations,  the  Laplace transforms of the memory functions
 are interrelated as follows 
\begin{equation} \label{Psi-Phi} 
 1 + \widetilde{\Psi} (s) = \frac{1}{1 - \widetilde{\Phi} (s)}\,. 
 \end{equation}

According to their  behaviour  for large times 
the viscoelastic models can be considered   solid-like
 (finite creep and relaxation)  and fluid-like (infinite creep and complete relaxation), see e.g. \cite{Pipkin_BOOK86}. 
As a consequence for fluid-like models we have      
\begin{equation} \label{type II}
 \J(+\infty) =+ \infty\,, \quad \G(+\infty) = 0\,. 
 \end{equation}
We note the the simplest model of this type  is the classical Maxwell body characterized by a single time constant $\tau_*$ 
\begin{equation}
 \Psi(t)  = \frac{1}{\tau_*}\,, \quad
\Phi(t) =    \frac{1}{\tau_*}\,  \exp (-t/\tau_*)\,.
\end{equation}
More generally, the fluid-like models exhibit continuous (or discrete) time spectra   that we write as
\begin{equation}\label{time-spectra}
\begin{split}
\Psi(t) &= a + b \int_0^\infty   H_{\Psi} (\tau)\,
\exp (-t/\tau)\, d\tau\,, \\
\Phi(t) &= c  \int_0^\infty   H_{\Phi} (\tau)\,
\exp(-t/\tau)\, d\tau\,, 
\end{split}
\end{equation}
with $a, b, c \ge 0$ 
and $H_{\Psi}(\tau)\ge 0$,  $H_{\Phi}(\tau) \ge 0$.

It is worth to remark that we can recover the memory functions for  the  Maxwell model by setting in Eqs. (\ref{time-spectra}):
\begin{equation}
\begin{split}
a &= \frac{1}{\tau_*}\,, \quad b  =0\,,\quad  H_{\Psi} (\tau) \equiv 0\,,\\ c&=\frac{1}{\tau_*}\,, \quad H_{\Phi}(\tau) = \delta(\tau-\tau_*)\,,
\end{split}
\end{equation}
where $\delta$ denotes the Dirac  generalized function.

Physically, the mathematical representations 
(\ref{time-spectra})  imply that
any   processes in linear viscoelasticity can be generally considered  as a continuous (or discrete) superposition of elementary (that is exponential) relaxation processes.    

\section{The Bessel models}
	In this section we discuss a generalization of the fluid-like model proposed by Giusti and Mainardi in \cite{AG-FM_MECC16} 
	based on modified Bessel functions of order $0,1,2$ in the Laplace domain. 
	Indeed, turning  to modified Bessel functions of contiguous order 
	$\nu>-1$,   we propose a suitable   generalization  of  the memory functions related to the rate of creep and relaxation in the Laplace domain.  Consequently, by inversion, we compute both memory functions   and  the corresponding material functions in the time domain  for this class of viscoelastic models that we have agreed to call
{\it Bessel models}. 	
%	Due to the relation of these viscoelastic models with the Bessel
 % % functions   we agree to refer them to as {\it Bessel models} 
\subsection{Creep and Relaxation rates}
	Consider a linear viscoelastic model characterized by  a \textit{rate of creep function} $\Psi _{\nu} (t)$  defined, in the Laplace domain, as
\begin{equation} \label{eq_LT_Psi}
	\widetilde{\Psi} _{\nu} (s) = \frac{2 (\nu + 1)}{\sqrt{s}} \frac{I_{\nu + 1} (\sqrt{s})}{I_{\nu + 2} (\sqrt{s})} \, ,
	 \quad \nu > -1 \,.
	\end{equation}
	Now, the Laplace transform of the  corresponding    \textit{rate of relaxation function}
	$\widetilde \Phi_\nu(t) $   is related to 
	$\widetilde\Psi_\nu(s)$ 
	 as follows
	\begin{equation} \label{eq_reciprocity}
	 1 + \widetilde{\Psi} _{\nu} (s) = \frac{1}{1 - \widetilde{\Phi} _{\nu} (s)} \,,
	\end{equation}
	as it is required by the general  relation (\ref{Psi-Phi}) between the two memory functions.
	
	Taking into account 
	the recurrence relations for the modified Bessel functions of the first kind $I_{\nu}$, \ie
\begin{equation}
 I_{\nu - 1} (z) - \frac{2 \nu}{z} I_{\nu} (z) = I_{\nu + 1} (z) \, , 
 \end{equation}
	we immediately get the expression for the Laplace transform of the \textit{relaxation memory function}\footnote{\blue{For sake of completeness, it is possible to give some functional bounds for these ratios of Bessel functions, in the Laplace domain (see e.g. \cite{extra}).}} $\Phi _{\nu} (t)$, \ie
	\begin{equation}\label{eq_LT_Phi}
	\widetilde{\Phi} _{\nu} (s) = \frac{2 (\nu + 1)}{\sqrt{s}} \frac{I_{\nu + 1} (\sqrt{s})}{I_{\nu} (\sqrt{s})} \, , \quad \nu > - 1\,.
	\end{equation}

	Let us first focus on the Laplace Transform of the \textit{Rate of Creep} $\widetilde{\Psi} _{\nu} (s)$ in Eq. (\ref{eq_LT_Psi}). 
We can obtain the Rate of Creep in the time domain by inverting the Laplace transform $\widetilde{\Psi} _{\nu} (s)$. To do this we apply the Bromwich theorem,
	\begin{equation}\label{Bromwich}
\Psi _{\nu} (t) = \frac{1}{2 \pi i} \int _{Br}  \widetilde{\Psi} _{\nu} (s) \, e^{st} \, ds\,,
\end{equation}
	where with $Br$ we denote the so called Bromwich contour.
	Now, it is easy to see  	from the power series of  
	Modified Bessel functions of the First Kind:
\begin{equation}\label{Eq_def_bessel}
I_{\nu} (z) = \left( \frac{z}{2} \right) ^{\nu} \, \sum _{k=0} ^{\infty} \frac{1}{k! \, \Gamma (\nu + k + 1)} \left( \frac{z}{2} \right) ^{2 k}
 \, ,
\end{equation}
	that the function $\widetilde{\Psi} _{\nu} (s)$ is single valued and has a simple pole in $s=0$.  Indeed
	\begin{equation}\label{Eq_s_zero}
	\begin{split}
	\widetilde{\Psi} _{\nu} (s) &= \frac{2 (\nu + 1)}{\sqrt{s}} \frac{I_{\nu + 1} (\sqrt{s})}{I_{\nu + 2} (\sqrt{s})} \simeq\\
	&\simeq \frac{4 (\nu + 1)(\nu + 2)}{s} \, , \qquad \mbox{as} \,\, s \to 0\,,
	\end{split}	
	\end{equation}
	where we have made use also of the known property of the Gamma function  $\Gamma (z + 1) = z \Gamma (z)$.
Furthermore, it is straightforward to see that $\widetilde{\Psi} _{\nu} (s)$ has  infinite simple poles in $s = - j_{\nu + 2 , \, n} ^{2}$, where $j_{\nu + 2 , \, n}$ represents the $n$th positive real root of the Bessel functions  of the corresponding order 
	$J _{\nu + 2} (x) = 0$ ($n = 1, 2, 3, \ldots$).  Indeed,
\begin{equation}
 I_{\nu+2} (\sqrt{s}) = 0 \quad \Longleftrightarrow \quad J_{\nu+2} (\lambda) = 0 \, ,
 \end{equation}
 having renamed  $\sqrt{s} = - i \lambda$ with $\lambda  >0$
 and recalled the known relation $I_{\nu} (iz) = i^{\nu} J_{\nu} (z)$.
%% 
 %  $J_\nu$ is the standard Bessel function of the first kind.	
Therefore, the non-vanishing simple poles for the Laplace transform of the rate of creep are given by
\begin{equation}
s_n = - j_{\nu + 2 , \, n} ^{2} \, , \quad n = 1, 2, \ldots
\end{equation}

Now, we can safely compute the Bromwich integral
(\ref{Bromwich})
 by means of the Residues theorem, \ie
\begin{equation}
\begin{split}
\Psi _{\nu} (t) &= \sum _{s_n} \res \left\{ \widetilde{\Psi} _{\nu} (s) \, e^{st} \, ; \, s = s_n \right\} =\\
&= \sum _{n = 0} ^\infty \res \left\{ \frac{2 (\nu + 1)}{\sqrt{s}} \, \frac{I_{\nu + 1} (\sqrt{s})}{I_{\nu + 2} (\sqrt{s})} \, e^{st} \, ; \, s = s_{n} \right\}
\end{split}
\end{equation}
where $s_{n} = - j_{\nu + 2 , \, n} ^{2} \neq 0$, for $n \geq 1$, and $s_{0} = 0$.
Thus, we have to distinguish two cases:\\
If $s_n \neq 0$, then
\begin{equation}
\res \left\{ \widetilde{\Psi} _{\nu} (s) \, e^{st} \, ; \, s = s_{n} \right\} = 4 (\nu + 1) \, \exp \left( s_n t \right) \, .
\end{equation}
This result is obtained after a few passages hereafter outlined.
\begin{equation}
\begin{split}
& \res \left\{ \frac{2(\nu+1)}{\sqrt{s}} 
\frac{I_{\nu + 1} (\sqrt{s}) }{I_{\nu+2} (\sqrt{s})} \, e^{st}\right\} _{s = s_{ n}} 
=\\
&= \lim_{s \to s_{n}}  (s - s_{ n}) \,
 \frac{2 (\nu + 1)}{\sqrt{s}} 
\frac{I_{\nu + 1} (\sqrt{s}) }{I_{\nu+2} (\sqrt{s})} \,e^{st} \\
&=  \lim_{s \to s_{n}} \, \frac{2 (\nu + 1)}{\sqrt{s}} 
\frac{I_{\nu + 1} (\sqrt{s})}
{I^\prime_{\nu+2} (\sqrt{s})/(2\sqrt{s})}\, e^{st}=\\  
&=4 (\nu +1) \, \exp {\left( s_{ n} t \right)}\, . 
\end{split} 
\end{equation}
Above  we have used the property that if $s_{n}$ is a zero of the modified Bessel function $ I_{\nu+2}$,  then we have,
\begin{equation}
I_{\nu+2}^\prime(s_{ n}) = I_{\nu+1} (s_{n}) \,,
\end{equation}
as we deduce from the general identity for modified Bessel functions, see e.g.  \cite{AS 1965}, 
$$ I'_\nu(z) 
= -\frac{\nu}{z} I_\nu(z) + I_{\nu-1} (z)
= \frac{\nu}{z} I_\nu(z) + I_{\nu+1} (z)\,.$$
Otherwise, if $s = s_{0} = 0$, in view of Eq.(\ref{Eq_s_zero})
\begin{equation}
\begin{split}
\res \left\{ \widetilde{\Psi} _{\nu} (s) \, e^{st} \, ; \, s=0 \right\} &= \lim _{s \to 0} s \, \widetilde{\Psi} _{\nu} (s) \, e^{st} =\\
&= 4 (\nu + 1) (\nu + 2) \,.
\end{split}
\end{equation}
	Hence, the memory function is given by
\begin{equation}
\begin{split}
	\Psi _{\nu} (t) = &4 (\nu + 1) (\nu + 2) +\\
	&+ 4 (\nu + 1) \sum_{n=1} ^\infty \exp \left( - j_{\nu + 2 , \, n} ^{2} \, t \right) \, , \quad \nu > -1\,, \label{Dirichlet_2}
\end{split}	
	\end{equation}
where $j_{\nu + 2 , \, n}$ is the $n$th positive real zero of $J _{\nu + 2}$ and $t>0$ is the time.
	
As a consequence,  it can be easily noticed that the time representation of the rate of creep $\Psi _{\nu} (t)$ in Eq. (\ref{Dirichlet_2}) is a \textit{Dirichlet series}. In particular, the series in Eq. (\ref{Dirichlet_2}) is absolutely convergent for $t>0$. However, we omit the details of this proof that can be easily deduced following the procedures displayed in \cite{AG-FM_MECC16} and 
	\cite{AG-FM_SNEDDON16}.
	
	Proceeding in strict analogy with the above discussion, one can easily find an explicit form for the relaxation memory function $\Phi _{\nu} (t)$. Indeed, we find that
\begin{equation}
\Phi _{\nu} (t) = 4 (\nu + 1) \, \sum_{n=1} ^\infty \exp \left( - j_{\nu , \, n} ^{2} \, t \right) \,,\label{Dirichlet_1}
\end{equation}
which is a Dirichlet series  absolutely convergent
 for $t>0$, where $j_{\nu , \, n}$ is the $n$th positive real zero of
  $J _{\nu}$ analogously with the discussion above for 
  $\Psi _{\nu} (t)$.

In the particular case  $\nu=0$  the  memory functions $\Psi_\nu$ and  $\Phi_\nu$ reduce to those  dealt in the previous paper   
\cite{AG-FM_MECC16}.
 
Comparing  Eqs. (\ref{Dirichlet_2}),  (\ref{Dirichlet_1}) with
Eqs. (\ref{time-spectra}) we note that   the time spectra of the two memory functions are discrete  with characteristic (non-dimensional) times $1/  j_{\nu + 2 , \, n}$, $1/j_{\nu, \, n}$.
 % Being the characteristic times related to the zeros of the Bessel
  % functions,  we agree to refer our models to as {\it Bessel models}  
%  Due to the relation of these viscoelastic models with the Bessel % % functions   we agree to refer them to as {\it Bessel models}.

	\subsection{Material functions}
	Following the discussion in Section \ref{section_2}, we have that
\begin{equation}
\begin{split}
& s \widetilde{\J}_{\nu} (s) = 1 + \widetilde{\Psi} _{\nu} (s) \,\, \div \,\, \J' _\nu(t) = \Psi _{\nu} (t) \, , \\
& s \widetilde{G}_\nu (s) = 1 - \widetilde{\Phi} _{\nu} (s) \,\, \div \,\, \G' _\nu(t) = - \Phi _{\nu} (t) \, ,
\end{split}
\end{equation}
by setting for convenience $\J_\nu (0^+) = \G_\nu(0+) = 1$.\\
Hence we get
\begin{equation} \label{J-G-integrals}
\begin{split}
\J_\nu(t) &= 1 + \int _0 ^t \Psi _{\nu} (u) \, du  \,, \\
\G_\nu(t) &= 1 - \int _0 ^t \Phi _{\nu} (u) \, du \, .
\end{split}
\end{equation}
%% As discussed in the literature (see \cite{AG-FM-2016}, \cite{AG-%% FM-2016}).
Integrating term by term the Dirichlet series defining 
$\Psi _{\nu} (t)$ and $\Phi _{\nu} (t)$ we get
\begin{equation} \label{eq_J1}
\begin{split}
 \J_\nu(t) &\!=\! 1 + 4 (\nu + 1) (\nu + 2) t + 4 (\nu + 1) \times\\
&\times \!\sum _{n=1} ^{\infty} \!\left[ \!\frac{1}{j_{\nu + 2 , \, n} ^{2}}
 - \!\frac{1}{j_{\nu + 2 , \, n} ^{2}} \exp \left( - j_{\nu + 2 , \, n} ^{2} \, t \right) \right] \, ,
 \end{split} 
\end{equation}
\begin{equation} \label{eq_G1}
\G_\nu(t) \!=\! 1 - \!4 (\nu + 1)\! \sum _{n=1} ^{\infty} \left[ \frac{1}{j_{\nu , \, n} ^{2}} - \frac{1}{j_{\nu , \, n} ^{2}} \exp \left( - j_{\nu , \, n} ^{2} \, t \right) \right] \, .
\end{equation}
Now, if we recall the following result concerning the zeros of the Bessel functions of the first kind (see \eg \cite{Sneddon 1960}, 
\cite{AG-FM_SNEDDON16})
\begin{equation} \label{SNEDDON}
\sum _{n=1} ^{\infty} \frac{1}{j_{\nu , \, n} ^{2}} = \frac{1}{4 (\nu + 1)}    \, ,
\end{equation}
then we get
\begin{equation} \label{eq_J2}
\begin{split}
\J_\nu(t) = &2 \left( \frac{\nu + 2}{\nu + 3} \right) + 4 (\nu + 1) (\nu + 2) t\\
&- 4 (\nu + 1) \sum _{n=1} ^{\infty} \frac{1}{j_{\nu + 2 , \, n} ^{2}} \exp \left( - j_{\nu + 2 , \, n} ^{2} \, t \right) \, , 
\end{split}
\end{equation}
\begin{equation} \label{eq_G2}
\G_\nu(t) = 4 (\nu + 1) \sum _{n=1} ^{\infty} \frac{1}{j_{\nu , \, n} ^{2}} \exp \left( - j_{\nu , \, n} ^{2} \, t \right) \, .
\end{equation}

\section{Numerical results and asymptotic representations}  \label{section_4}
For our Bessel models we now exhibit versus non-dimensional time the plots  of the creep and relaxation  memory functions $\Psi_\nu(t)$, $\Phi_\nu(t)$ according to Eqs. (\ref{Dirichlet_2}),  (\ref{Dirichlet_1}), and of the corresponding material functions $\J_\nu(t)$ and 
$\G_\nu(t)$ according to Eqs. (\ref{eq_J2}),  (\ref{eq_G2}).
The  corresponding Dirichlet series have been computed by taking 100 terms  to ensure a suitable numerical convergence. We have used   an efficient  MATLAB routine given in \cite{MATLAB} to compute the requested zeros of the Bessel functions. In order to have an idea of the effect of the parameter $\nu>-1$ we have compared the plots  selecting   the values
 $\nu= -0.5, 0, 0.5, 1,$ and using suitable logarithmic scales.
%%%%%%%%
\begin{center}
\includegraphics[width=9cm]{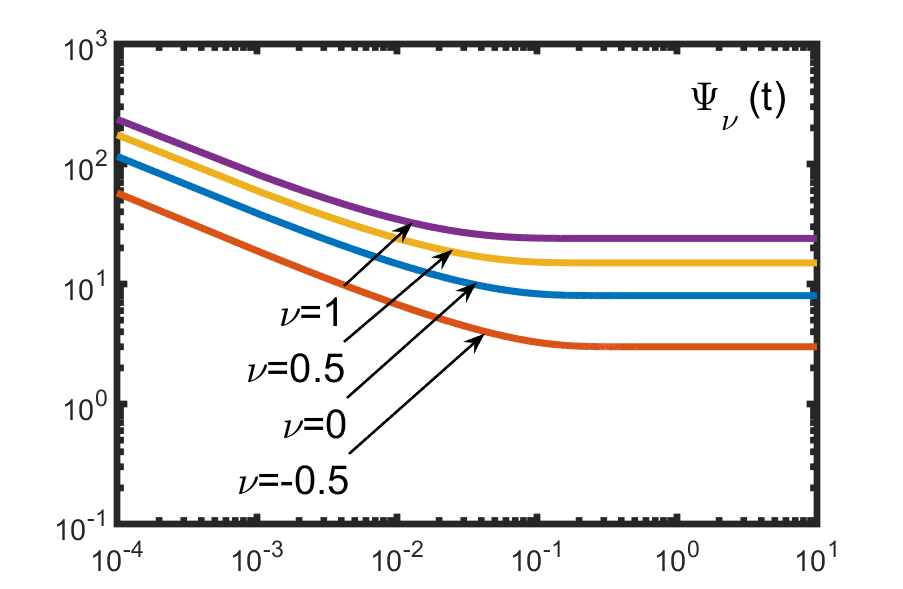}
\end{center}
Fig.1: The creep-memory  function $\Psi_\nu(t)$  for $\nu=-0.5, 0, 0.5, 1.$ 
\begin{center}
	\includegraphics[width=9cm]{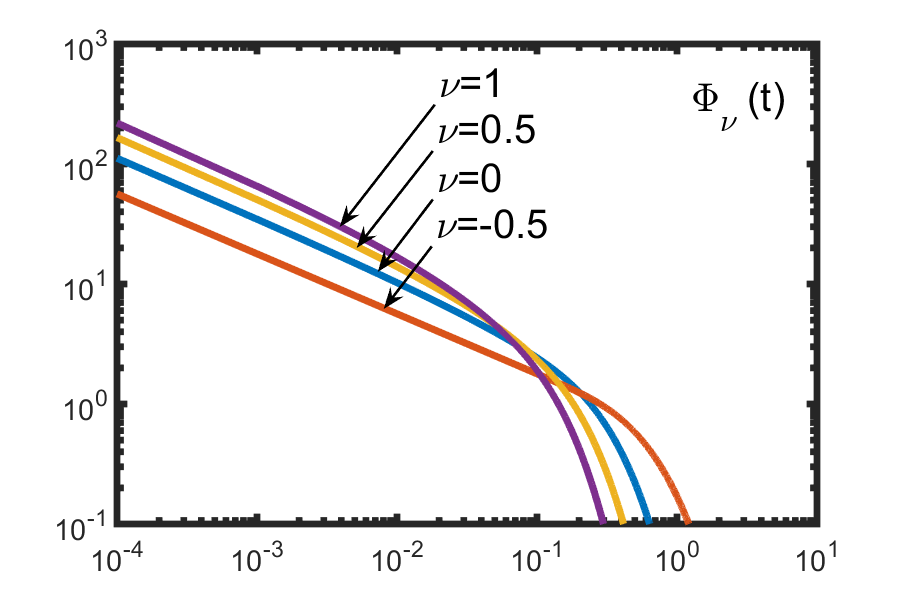}
\end{center}
Fig.2: The relaxation-memory  function $\Phi_\nu(t)$  for $\nu=-0.5, 0, 0.5, 1.$
\begin{center}
	\includegraphics[width=9cm]{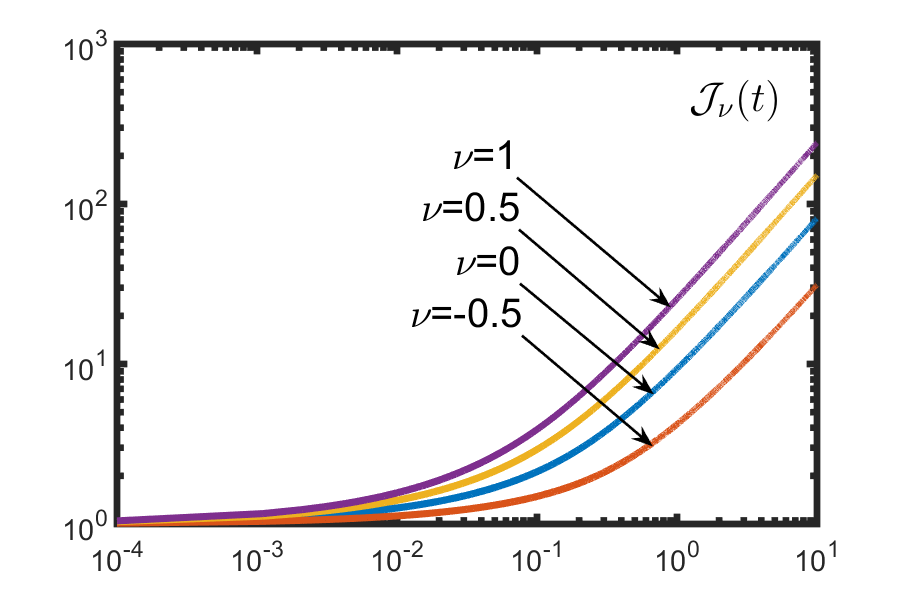}
\end{center}
Fig.3: The creep compliance  $\J_\nu(t)$  for $\nu=-0.5, 0, 0.5, 1.$
\begin{center}
	\includegraphics[width=9cm]{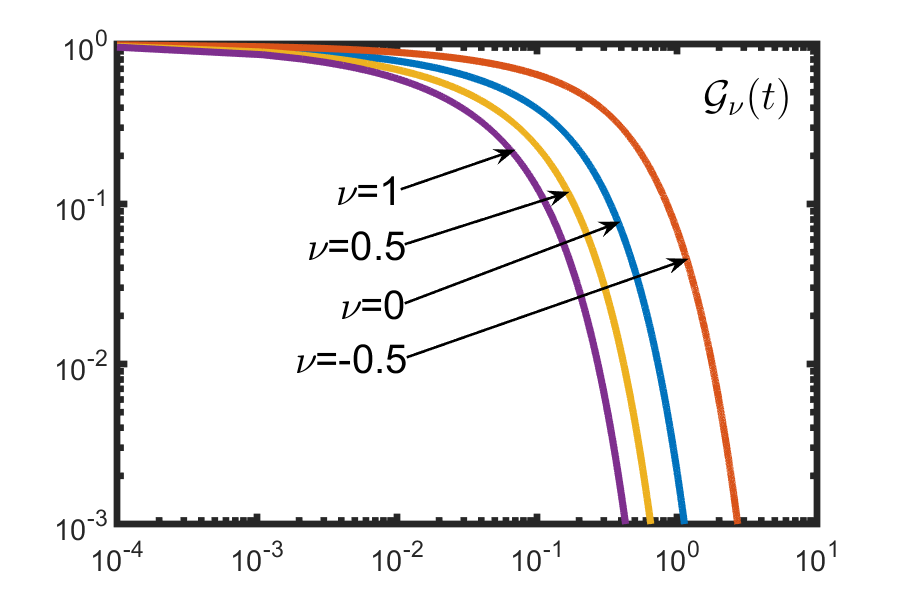}
\end{center}
Fig.4: The relaxation modulus $\G_\nu(t)$  for $\nu=-0.5, 0,0.5, 1.$
%%%%%%%
\vskip 0.5truecm
We are also interested to see how the asymptotic representations of the memory and material functions for short and long time are matching with the corresponding numerical solutions.
For sake of brevity we limit ourselves to consider the case 
$\nu=1$ that would provide a sufficient overview of the matching. 
We remark that the case $\nu=0$ was already considered in the recent paper by Giusti and Mainardi \cite{AG-FM_MECC16}.   

The asymptotic representations  of the memory  functions  for short  times are {\it formally} derived 
according  to a Tauberian theorem
by inverting their  Laplace transforms  
(\ref{eq_LT_Psi}), (\ref{eq_LT_Phi})
	approximated as $s\to \infty$. 
	For this purpose we use
the known asymptotic representation of the generic modified Bessel function 
\begin{equation}
I_\nu(z) \sim \frac{e^z}{\sqrt{2\pi}\, z^{1/2}}\,, 
\quad  |z| \to \infty, \; |\hbox{arg} \,z| <\frac{\pi}{2}\,.
\end{equation}    
For long times it is sufficient to take the first term of the corresponding expressions (\ref{Dirichlet_2}),  (\ref{Dirichlet_1}). 
Finally  the asymptotic representations read:
\begin{equation} \label{eq_Psi-as}
\Psi_\nu (t) 
\sim \left \{ 
\begin{array}{ll}
{\ds \frac{2(\nu+1)}{\sqrt{\pi}} \, t^{-1/2} \,, }\quad &  t \to 0\,, 
\\ 
{\ds 4(\nu+1)(\nu+2)   \,, }\quad &  t \to \infty\,, 
\end{array}
\right.
\end{equation}
and
\begin{equation} \label{eq_Phi-as}
\Phi_\nu (t) 
\sim \left \{ 
\begin{array}{ll}
{\ds \frac{2(\nu+1)}{\sqrt{\pi}} \, t^{-1/2}\,,} \quad &  t \to 0\,, 
\\
{\ds 4(\nu+1) \, \exp (- j_{\nu,1}^2 t ) \,,} \quad &  t \to \infty \,,  
\end{array} \right.
\end{equation}

  In the subsequent Figures 5-6-7-8. we show in   continuous lines the plots of the memory and material functions  versus time,  in comparison   with the corresponding asymptotic representations for short and long times, printed with  dashed and dotted lines, respectively. 
 %\begin{figure}
% \centering
\begin{center}
\includegraphics[width=9cm]{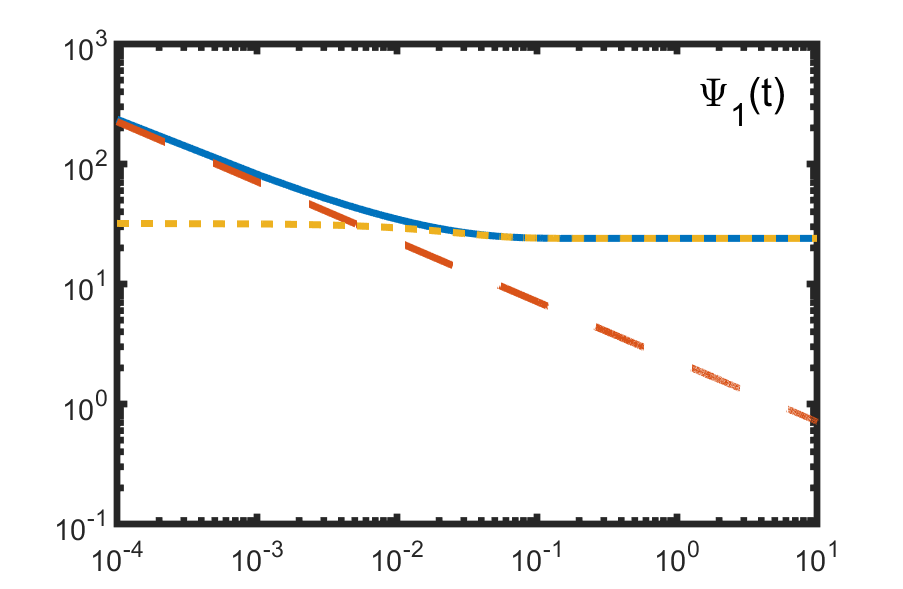}
\end{center}
Fig.5: The  memory  function $\Psi_1(t)$ 
in comparison with its asymptotic approximations.
%(continuous line) with its asymptotic representations (dotted lines)
% \end{figure}
\begin{center}
\includegraphics[width=9cm]{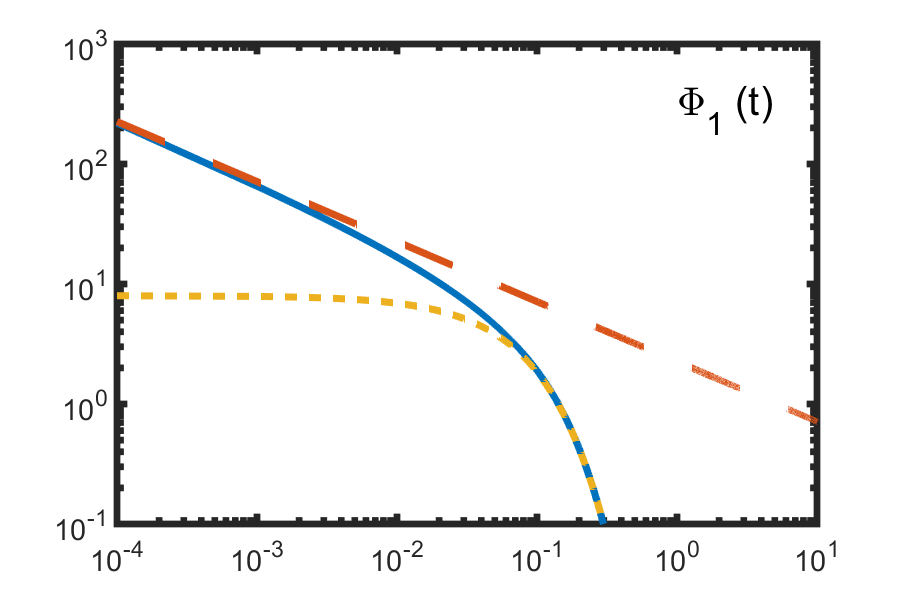}
\end{center}
Fig.6: The  memory  function $\Phi_1(t)$
in comparison with its asymptotic approximations. 
%(continuous line) with its asymptotic representations (dotted lines) 
% \end{figure}
\begin{center}
\includegraphics[width=9cm]{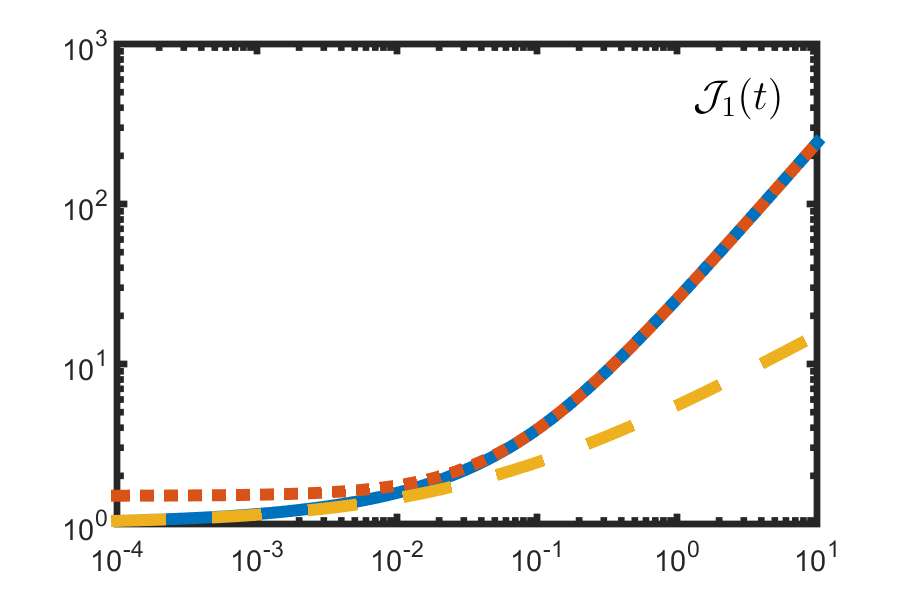}
\end{center}
Fig.7: The creep   compliance $\J_1(t)$ 
in comparison with its asymptotic approximations.
% (continuous line) with its asymptotic representations (dotted lines) v
% \end{figure}
\begin{center}
\includegraphics[width=9cm]{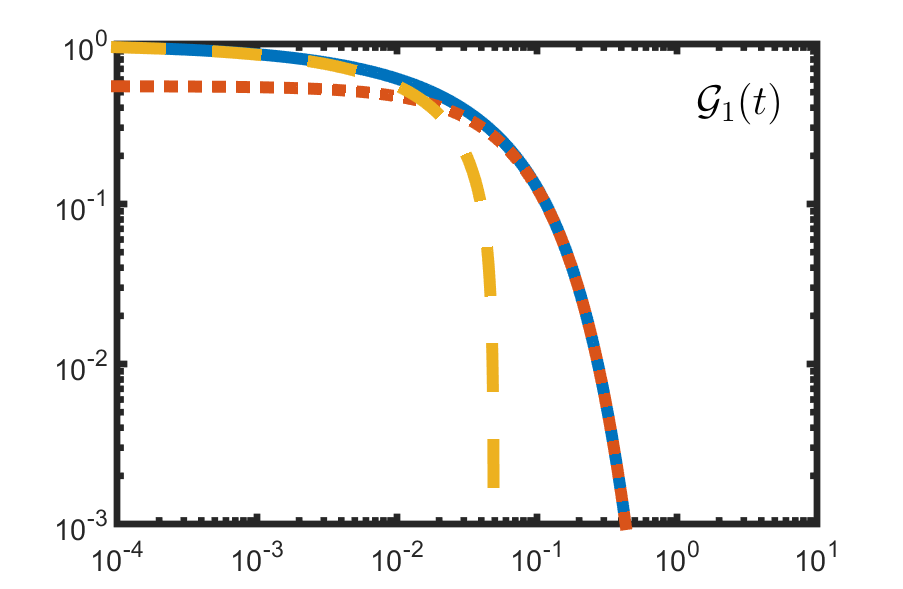}
\end{center}
Fig.8: The relaxation modulus $\G_1(t)$ 
in comparison with its asymptotic approximations.
%(continuous line) with its asymptotic representations (dotted lines) 
% \end{figure}

\vskip 0.25truecm
The asymptotic representations of  $\J_\nu (t)$ and $\G_\nu (t)$ are easily derived  from  Eqs. (\ref{J-G-integrals})
by integrating 
$\Psi_\nu (t)$  and $\Phi_\nu (t)$ 
given by  (\ref{eq_Psi-as}),  (\ref{eq_Phi-as}), respectively. 
They read:
\begin{equation} \label{eq_J-as}
\J _\nu (t)
\sim \left \{ \begin{array}{ll}
 1+{\ds  \frac{4(\nu+1)}{\sqrt{\pi}} }\, t^{1/2}\,,  &  t \to 0, 
 \\ 
2{\ds \frac{(\nu+2)}{(\nu+3)} }+ 4(\nu+1)(\nu+2)t\, , &t \to \infty,
\end{array}
\right.
\end{equation}
and
\begin{equation} \label{eq_G-as}
\G_\nu (t) 
\sim \left \{ \begin{array}{ll}
 1- {\ds \frac{4(\nu+1)}{\sqrt{\pi}}} \, t^{1/2} \,, &  t \to 0\,, 
 \\ 
4(\nu+1){\ds \frac{1}{j_{\nu,1}^2}}\,\exp (- j_{\nu,1}^2 t )\,, 
& t \to \infty\,, 
\end{array} \right.
\end{equation}

It is  interesting to show in Table I  the numerical values of the first positive zero $j_{\nu, 1}$ of the Bessel function $J_\nu$  and of its square for the 4 cases of $\nu$ considered in the plots.  
This way we get an hint on how the exponential decay in relaxation behaves as $\nu$ increases, see Eqs. (\ref{eq_Phi-as}),
(\ref{eq_G-as}). 

\vskip 0.25truecm
\begin{center}
\begin{tabular}{|c|c|c|c|c|}
\hline
$\nu$ & -0.5 & 0 & 0.5 & 1 \\
\hline
$j_{\nu,1}$ & 1.57 & 2.40 & 3.14 & 3.83 \\
\hline
$j_{\nu,1}^2$ & 2.47 & 5.78 & 9.87 & 14.68 \\
\hline
\end{tabular}
\vskip 0.25truecm
Table I
 \end{center}

Based on the above asymptotic  results and on the theory of fractional viscoelastic models dealt with detail in \cite{Mainardi_BOOK10,Mainardi-Spada 2011}, 
 we recognize that the creep and relaxation properties of the  viscoelastic models  under consideration are compatible with those of a peculiar model.  In particular, they are consistent with a fluid-like body, that, for short times, is akin with a fractional Maxwell model of order 1/2 and, for large times, is similar to a standard Maxwell model.

\section{Conclusions and final remarks}
	In this paper we have illustrated a parametric class of linear viscoelastic models of fluid-like type
depending on a parameter $\nu>-1$.  In particular, for $\nu=0$ we recover  the model recently introduced by Giusti and Mainardi \cite{AG-FM_MECC16} for describing the effect of fluid viscosity for  fluid filled  elastic tubes.
 In the Laplace domain the memory and material  functions characterizing this class  are represented by ratios of modified Bessel functions of contiguous order $\nu$, so that we agree to refer them to as {\it Bessel models}.    
	In analogy with the case $\nu=0$,  the models of this general class exhibit (with a different strength and time range depending on $\nu$) the  creep and relaxation properties typical of a fractional Maxwell body of order 1/2  for short times, and of a standard Maxwell body, for long times.  
 We have noted from the plots in Figs 1-4  that the time range of validity of short time approximation 
  is shorter as higher is the value of the parameter 
 $\nu$.  This effect appears clear  from inclined (parallel) straight lines in log-log scales. 
 
 However, this asymptotic behavior of power law type with exponent $\pm 1/2$  exhibited  by the memory and material functions,   even though for a restricted range of time after the initial  instant $t=0$,
 is peculiar of the whole class of  Bessel models.   
  
 The above  analysis  is worth to be pursued  in future with respect 
%% to its realizability with an equivalent circuit and 
to the evolution of the transient waves propagating through
viscoelastic media  modelled by this class. 
Indeed, as formerly pointed out in 1975 by 
Buchen and Mainardi in \cite{Buchen-Mainardi 1975} (see also \cite{Mainardi_BOOK10}), the singular behavior of the memory and material  functions   at $t=0^+$    
induces a wave-front smoothing of any initial discontinuity, akin to  a diffusion effect.
For more details on  this smoothing effect found in  transient waves propagating in  media  with  singular memory, we refer to Hanyga \cite{Hanyga 2001}.

	\blue{It is also interesting to stress that such models could be of a certain interests for the study of physical properties of polymers. Indeed, this class of models could turn out to be useful to describe polymers featuring dynamical behavior akin to the one described above. In such cases, the value for $\nu$ could then be appropriately tuned in order to fit the experimental data.}

\section*{Acknowledgements}
The work of A. G. and F. M. has been carried out in the framework of the activities of the National Group of Mathematical Physics (GNFM, INdAM).

%%%%%% Bibliography


\begin{thebibliography}{}
    
\bibitem{AS 1965}
M.~Abramowitz and I.A.~Stegun,
{\it Handbook of Mathematical Functions}, Dover, New York (1965).

\bibitem{Buchen-Mainardi 1975}
P.W. Buchen, F. Mainardi,
Asymptotic expansions for transient viscoelastic waves,
{\it Journal de M{\'e}canique} {14},  597--608 (1975).

\bibitem{AG-FM_MECC16}
A. Giusti, F. Mainardi, 
A dynamic viscoelastic analogy for fluid-filled elastic tubes,
{\it Meccanica}, published on line, 04 February  2016.
DOI: 10.1007/s11012-016-0376-4 
{\em E-print}, 
\href{http://arxiv.org/abs/1505.06694}{arXiv:1505.06695} (2015).

\bibitem{AG-FM_SNEDDON16}
A. Giusti, F. Mainardi, 
On infinite series concerning zeros of Bessel functions of the first kind, 
{\em E-print}, 
\href{http://arxiv.org/abs/1601.00563}{arXiv:1601.00563} (2016).

\bibitem{Hanyga 2001}
A. Hanyga, 
Wave propagation in media with singular memory,
{\it Mathematical and Computer Modelling} {34},  1399--1421 (2001).

\bibitem{Hanyga STAMM04}  %% REVISED-NEW
A. Hanyga,  
Physically acceptable viscoelastic models,
in    Hutter, K. and Wang, Y. (Editors),
{\it Trends in Applications of Mathematics to Mechanics},
%% Proceedings {\it  Workshop STAMM'04, Seheim (Germany) 22-28 August 2004},
Shaker Verlag GmbH, Aachen, pp. 125--136 (2005).
%[{\tt www.geo.uib.no/hjemmesider/andrzej/index.html}]


\bibitem{Hardy-Riesz 1915}
G.H. Hardy, M. Riesz, 
{\it The General Theory of Dirichlet Series},
Cambridge University Press, Cambridge (1915).

\bibitem{extra}
\blue{A. Laforgia, P. Natalini, 
Some Inequalities for Modified Bessel Functions, 
{\it Journal of Inequalities and Applications}, Volume 2010, 
Article ID 253035, 10 pages.}

\bibitem{Mainardi_BOOK10}
 F.~Mainardi, 
     {\it Fractional Calculus and Waves in Linear Viscoelasticity},
     Imperial College Press,  London (2010). 

\bibitem{Mainardi-Spada 2011}
F. ~Mainardi, G. ~Spada,
Creep, relaxation and viscosity properties for basic fractional models in rheology,
{\it Eur. Phys. J. Special Topics} {193}, 133--160 (2011).
{\em E-print}, 
\href{http://arxiv.org/abs/1110.3400}{arXiv:1110.3400} (2011).

\bibitem{MATLAB}
MATLAB Central, File Exchange,
{\it Bessel Function Zeros} by Greg von Winckel, File ID: \#6794 (Updated 27 Jan 2005)
%{\it Bessel Zero Solver} by Jason Nicholson,
% File ID: 48403 (Updated 10 Nov 2014).

\bibitem{Pipkin_BOOK86}
A.C. Pipkin,
{\it Lectures on Viscoelastic Theory},
Springer-Verlag, New York (1986).

\bibitem{Schilling-et-al 2012}   %% NEW 
R.L. Schilling, R.  Song,   Z.  Vondracek,  
{\it Bernstein Functions. Theory and Applications}, 
2-nd Edition, De Gruyter, Berlin (2012).  %%  2012.

\bibitem{Sneddon 1960}
I. N. Sneddon,
On some infinite series involving the zeros of Bessel functions of 
the first kind,
{\it Proc. Glasgow Math. Assoc.} {4}, 144-156 (1960).

\end{thebibliography}
\end{document}